# A Survey and Taxonomy of Graph Sampling

Pili Hu and Wing Cheong Lau

Department of Information Engineering
Chinese University of Hong Kong
{hupili,wclau}@ie.cuhk.edu.hk

## Abstract

Graph sampling is a technique to pick a subset of vertices and/ or edges from original graph. It has a wide spectrum of applications, e.g. survey hidden population in sociology [54], visualize social graph [29], scale down Internet AS graph [27], graph sparsification [8], etc.

In some scenarios, the whole graph is known and the purpose of sampling is to obtain a smaller graph. In other scenarios, the graph is unknown and sampling is regarded as a way to explore the graph. Commonly used techniques are Vertex Sampling, Edge Sampling and Traversal Based Sampling. We provide a taxonomy of different graph sampling objectives and graph sampling approaches. The relations between these approaches are formally argued and a general framework to bridge theoretical analysis and practical implementation is provided.

Although being smaller in size, sampled graphs may be similar to original graphs in some way. We are particularly interested in what graph properties are preserved given a sampling procedure. If some properties are preserved, we can estimate them on the sampled graphs, which gives a way to construct efficient estimators. If one algorithm relies on the perserved properties, we can expect that it gives similar output on original and sampled graphs. This leads to a systematic way to accelerate a class of graph algorithms. In this survey, we discuss both classical text-book type properties and some advanced properties. The landscape is tabularized and we see a lot of missing works in this field. Some theoretical studies are collected in this survey and simple extensions are made. Most previous numerical evaluation works come in an ad hoc fashion, i.e. evaluate different type of graphs, different set of properties, and different sampling algorithms. A systematical and neutral evaluation is needed to shed light on further graph sampling studies.

# Table of contents

# 1 Introduction

People have defined many properties to characterize a graph, e.g. degree distribution, spectrum, NCut, etc. Those properties are very important for people to understand a network. They may also be further developed into criteria or objectives for some problems. It is interesting to know whether those properties are preserved on a transformed graph. If so, running the algorithm on the transformed graph has approximately the same effect as running it on original graph. Besides, we can estimate the properties of original graph using the transformed graph. A simple but effective way of graph transformation is via sampling: Select a subset of vertices or edges of the original graph. The biggest advantage of sampling methods are their execution efficiency so that the graph transformation procedure won't take longer time than straightforward computation on original graph.

## 1.1 Some Motivating Examples

In this section, we discuss some concrete examples and motivate graph sampling from different aspects:

- Lack of data. For example, you can not afford to crawl all the people from an Online Social Network (e.g. due to API call limit). Instead, you randomly pick IDs (it's sometimes possible to enumerate IDs for most OSNs) and crawl them. This is essentially sampling of vertices (edges are lost in a bulk). It is good to know how well this process preserves certain graph properties. Or more simple questions to ask is: how many vertices/ edges should we sample in order to get good coverage of certain information?

- Survey hidden population. In sociology studies, we may want to reach a hidden population, e.g. drug abusers. It is generally impossible to directly enumerate and sample from the whole population. Researchers usually start from a small set of seed nodes and expand according to their knowledge. Classical methods ranges from Snow Ball Sampling [17] to Respondent Driven Sampling [54].

- Graph Sparsification. Many modern complex networks are very large in size, making it hard to manipulate the whole graph. This calls for graph sparsification, which includes edge sparsification [8][32, L12] and vertex sparsification [44]. Graph sparsification is a classical problem. It often imposes very stringent mathematical constraints on the transformation, like preserving all cuts (cut sparsifiers) or perserving all quadratic forms of Laplacian (spectral sparsifiers). However, some of those techniques are too heavy to apply. We are interested in those as simple as straight sampling. See [21] for a survey.

- Reduce test cost. For example, Protein Interaction Network [60] is a frequent object to study in biochemical researches. The accurate test of the interactions between all possible neighbours may be too costly. In this case, one want to sample the graph and only test the edges on the sampled graph. Although there are more mature approaches (e.g. SPCA [70] for gene micro-arraying) to prioritize what edges should be tested first, sampling usually gives an easy first solution.

- Visualization. The original graph may be too big to fit in a screen. Displaying all the edges may be too cluttered. Sampling can give the graph a digest, which makes it easier for visualization [29]. Towards this end, we may hope the sampled graph "looks like" the original graph. This topic is also related to *Dimensionality Reduction* and *Graph Embedding*.

All those examples, although different in root motivation and mathematical depth, share the following common characteristics:

- The graph size is reduced during the transformation.

- Certain properties are expected from the output e.g. "being representative", "looking similar to original graph", etc.

## 1.2 Sampling as a Base Approach

In order to reduce the graph size and preserve graph properties at the same time, many complex approaches are possible. For example, one can formulate a mathematical programming problem in order to minimize the distance between original and sampled graph. This approach, while being very rigorous, can be extremely costly. For example, solving the graph cut sparsifier problem (reduce the graph size such that cuts are preserved) is NP-Hard [21]. Even when those polynomial approximation algorithms are used, they are still too complex. One may find that running the sparsifier is already more complex than running an algorithm on original graph. This deviates form our primary pursuit.

Another drawback of those complex methods is that they usually require global knowledge (the whole graph) in order to (approximately) solve the optimization. This makes it useless in some scenarios, e.g. Decentralized Social Networks (DSN), where we can only get part of the data and assume it is a sample (by some distribution) from the original graph.

With these observations, we are more interested simple techniques and see what result can be obtained. Graph sampling is a simple and intuitive approach. It is usually one of the first thoughts to handle massive data. It has been widely used but the performance is not systematically studied.

## 1.3 Organization

In Section 2, we introduce common notations used in this survey. In Section 3, we provide a taxonomy of graph sampling from three angles: objective, graph type and sampling approach. Relations between different objectives and different sampling approaches are discussed. Among common sampling approaches, Traversal Based Sampling (TBS) is a large class of algorithms and draws a lot research interest over the years. Towards this end, we devote the whole Section 4 to it. In Section 5, we discuss graph properties ranging from classical text-book type ones to advanced ones which may be more useful to support graph algorithms. Past works are collected and tabularized in this section, showing that a lot of future works can be done to numerically and theoretically study those properties. In Section 6, we collect some ad hoc theoretical results and make extensions where affordable in a short period of time. Lastly, we conclude the survey and discuss future works in Section 7.

# 2 Common Notations and Definitions

In this section, we provide some common definitions for this paper. Because there are a wide spectrum of surveyed works, many other symbols will be used (may be overrided) locally. We'll give definitions in the context.

We consider an unweighted and undirected graph, which will be the main object discussed in this paper. Without special mention, we call an unweighted and undirected graph as "graph" for short. Denote a graph as $G = <V, E>$. $V$ is the set of vertice and $E \subseteq \{(u,v) | u \in V, v \in V\}$ is the set of edges, where $(u,v)$ is an **unordered** pair. For convenience of discussion, we denote $n = |V|$ and $m = |E|$. The neighbourhood of vertex $v$ is $N(v) = \{u | (v,u) \in E, u \in V\}$. The degree of a vertex is defined as $d_G(v) = d(v) = d_v = |N(v)|$. These notations will be used interchangeably where appropriate. The vertices are denoted as $v_1, v_2, ... v_n$ or simply $1, 2, ... n$ as a shorthand. To facilitate our discussion, we define the incident edges to a vertex by $\delta(v) = \{(u,v) \in E | u \in N(v)\}$. The incident edges to a set of vertices $S$ is $\text{vol}(S) = \bigcup_{v \in S} \delta(v)$. The edges at the boundary of $S$ is $\delta(S) = \{(u,v) \in E | u \in S, v \notin S\}$. Note that $\delta(v) = \delta(\{v\}) = \text{vol}(v)$.

We briefly mention the definition for weighted or directed graphs for completeness.

- The weighted graph is $G = <V, E, w>$, where $w_e = w_{(u,v)} \in \mathbb{R}, \forall e = (u,v) \in E$.

- The directed graph is defined by changing the definition for edge from $(u,v)$ to $<u,v>$, where $<u,v>$ is an **ordered** pair.

Other symbols on those graphs can be defined accordingly.

We denote the sampled graph by $G_s = <V_s, E_s>$, where some validity conditions are required:

$$\begin{aligned} V_s &\in V \\ E_s &\in E \\ E_s &\subseteq \{(u,v) | u \in V_s, v \in V_s\} \end{aligned} \quad (1)$$

The first and second conditions ensure that the elements (either vertices or edges) are a sample from original graph. The third condition ensures the sampled elements compose to a valid graph. We denote $n_s = |V_s|$ and $m_s = |E_s|$, respectively.

The sampling procedure involves a primitive – "probing", e.g. use one API call to get someone's buddy list on an OSN, or asking previous participants in an experiment to refer some of their acquaintances. The budget is denoted as $B$. There is usually a unit cost $b$ associate with each probing operation. The cost and the information obtained during one probing vary according to the concrete application scenario.

Given a graph $G$, a property is defined as a function $f(G)$, where $f$ is probably a vector function in many cases. Note that this definition is different from the classical discussion of random graphs (e.g. [10] and [33] L5), where it is defined as a subset of graphs from the family, i.e. $Q \subseteq \mathcal{G}$. If $G$ is an r.v., so is $f(G)$. The classical definition is just an indicator for a certain $f(G)$. For example, define $Q$ as "subset of graphs who are connected"; Define $f(G): G \to \#$ of connected components; Obviously $G \in Q \Leftrightarrow I[f(G) = 1]$. Towards this end, we'll use the former definition, which is more general and implicitly adopted in a wide spectrum of works.

## 3 A Taxonomy of Sampling

There are many related works to "graph sampling". All these works have a sense of randomly picking vertices or edges (maybe according to current knowledge of the graph). However, they arise from different context and have different problem dimensions. In this section, we provide a taxonomy of surveyed graph sampling works.

### 3.1 Categorization By Sampling Objective

#### 3.1.1 Common Sampling Objectives

Common objectives of graph sampling are listed below:

1. Get representative subset of vertices. This is the usual motivation from sociology studies, e.g. poll the opinion of this sampled vertices (people). In many scenarios, target population can be sampled directly, e.g. phone number, random street survey, etc. In other scenarios, target population is hidden, e.g. drug abusers in urban area. In this latter case, the researchers have to execute certain sampling algorithms on a graph to explore the hidden population.

2. Preserve certain property of original graph. A property of a graph can be viewed as a (can be vector) function $f(G)$. Sometimes, we pursue exact property preservation (e.g. Section 6.4). Sometimes, we only preserve the property within certain error margin (e.g. Section 6.5). After performing sampling, two things can be done:

    - Estimate graph properties. If we know some property is perserved on $G_s$, we can calculate $f(G_s)$ as an estimator for $f(G)$. For example, MHRW (Section 4.4) preserves vertex label distributions (Section 5.1). So we can use $f(G_s)$ as an estimator for $f(G)$ if $G_s$ is obtain by MHRW from $G$ and $f(\cdot)$ is a vertex label distribution property. Note that for mere estimation purposes, one do not have to preserve the properties on $G_s$. As long as we know how the properties are biased, we may be able to correct it. RDS (Section 4.9) is such an approach.

    - Support graph algorithms. Many graph algorithms aim at optimizing certain objective associated with some graph properties. If we can preserve those properties on $G_s$, we may expect to obtain similar results by running the algorithm on $G_s$ instead of $G$. Many sampling approaches are very efficient (e.g. ES, VS, ESC, VSC; Section 3.3). This gives a general method to accelerate a class of graph algorithms.

3. Generate random graph. Graph generation is a big topic in its own right. However, some literature on graph generation also use the phrase like "graph sampling". There are some relations between the two topics. One can view a graph generation model as a family of graphs $\mathcal{G}$. The generation procedure is to sample one graph $G$ from $\mathcal{G}$. In our terminology, let $G = K^n$ be a complete graph of $n$ vertices. The process of performing an edge sampling (Section 3.3) on $G$ is indeed the generation of an Erdos-Renyi Network (ERN).

In this paper, we will focus on the second type, namely, property preservation. Both property estimation and algorithm supporting works are discussed. Note that different algorithms depend on graph properties in different ways, so the sampling procedures are often tailored to those algorithms. Towards this end, we also term the latter type of works as Problem Oriented Property Preservation (POPP).

### 3.1.2 Relations Between Property Preservation and Estimation

Although the motivation of property estimation and property preservation look different initially, they are closely related and can be sometimes transformed to each other. Consider the original graph $G$ and sampled graph $G_s$. Use cut weight as an example property:

- If we know cuts are preserved on $G_s$, we can directly compute the weight of the cut that we are interested in and it is naturally an estimator for it in $G$.

- If we know cuts are not preserved on $G_s$ but we have an estimator for the cuts, we can try to modify $G_s$ so that cuts are preserved. For example, if the edges are sampled with probability $p$, there are less edges in the cuts of $G_s$. However, we can upweight all the edges in $G_s$ by a factor of $1/p$ and the (weighted) cuts are then preserved. (e.g. Section 6.5.1)

We should note that the relationship is not symmetric. Namely, property preservation is actually a stronger and more useful notion. Here are a few remarks:

- All property preservation results can lead to property estimators.

- Not all property estimation results can be easily casted to property preservation results.

- In order to construct better estimators, people usually approach property estimation directly without first deriving a property preservation result. This simplifies the analysis because property preservation results are usually too strong for mere estimation purpose. In this way, property estimation is still of its own research interest.

In Section 6, we will discuss some ad hoc analytical results for both property estimation and property preservation.

## 3.2 Categorization By Type of Networks

Most works surveyed in this paper does not assume a specific type of network. The evaluation is also done on real-life social networks [22], online social networks [15], biological networks [18], Internet AS networks [27], P2P networks [50], etc. Lacking of an underlying graph generation model makes the theoretical analysis difficult in general. Towards this end, most works accompanied with theoretical analysis assumes certain type of graph generation model:

- Erdos-Renyi Network (ERN), also called "random graph", exponential random graph, Poisson random graph, etc. This is the most classical and well-understood probabilistic network model. See [10] for a comprehensive discussion of the properties of ERN.

- Power-Law Network (PLN), also called scale-free network. It was shown in the last decade that many networks present power-law degree distribution, ranging from social networks to biological networks. Works like [35] and [60] focus on this type of network.

- Small-World Network (SMN) [65] is a combination of ERN and regular graph. Many real social networks also exhibit some SMN properties. One notable property is the existence of efficient decentralized routing scheme [26].

- Fixed Degree Distribution Random Graph (FDDRG). In order to theoretically characterize the bias of BFS for degree distribution, [30] assumes a random graph model with fixed degree distribution. FDDRG is also termed as "configuration model" in [61].

## 3.3 Categorization By Sampling Approach

In this section, we first present most frequently discussed sampling approaches. Vertex Sampling and Edge Sampling are two classical sampling methods. They are also building blocks for more complex methods. Vertex Sampling with Neighbourhood operates like Vertex Sampling but neighbourhood information is acquired in one probing. We also discuss two variations to VS and ES with contraction. Traversal Based Sampling (TBS) is a large class of methods so we briefly mention it here and leave the full discussion to another dedicated section.

In the second part, we briefly discuss the relations between different sampling approaches and formalize the methodology for theoretical analysis and practical implementation.

### 3.3.1 Common Sampling Approaches

The commonly studied sampling techniques are:

- Vertex Sampling (VS). We first select $V_s \subseteq V$ directly without topology information (e.g. uniformly or according to some distribution on $V$). Then we let $E_s = \{(u,v) \in E | u \in V_s, v \in V_s\}$, namely only edges between sampled vertices are kept. One variation, stratified sampling, is commonly used in survey studies. Usually, only information associated with vertices is used, e.g. demographic attributes. We also regard it as vertex sampling.

- Edge Sampling (ES). We first select $E_s \subseteq E$ somehow. Then we let $V_s^{(1)} = \{u,v | (u,v) \in E_s\}$. This definition only arises in some theoretical discussions of basic sampling methods on graphs. A more realistic definition is to let $V_s^{(2)} = V$. Then the setting is same as graph (edge) sparsification. Graph sparsification has more complex quality metrics and algorithms. Edge sampling is just one of them. We do not deliberately distinguish the two definitions and one can find which one is used from context.

- Vertex Sampling with Neighbourhood (VSN). We first select $\tilde{V}_s \subseteq V$ directly without topology information. Then we let $E_s = \bigcup_{v \in \tilde{V}_s} \delta(v)$, and $V_s = \{u,v | (u,v) \in E_s\}$. We return $G_s = <V_s, E_s>$ as the sampled graph. Note that this is a more realistic setup than VS in social network crawling, where sampling a vertex means getting its buddy list. $|\tilde{V}_s|$ is limited by available resources (e.g. API call).

- Traversal Based Sampling (TBS) is a class of sampling methods. The sampler starts with a set of initial vertices (and/ or edges) and expand the sample based on current observations. This class of approaches arise naturally in the context of network crawling, hidden population survey, etc. Examples are like Snowball Sampling [17], Respondent Driven Sampling [22], Forest Fire [37], etc. It has a very long history and is also the recent research focus of these sampling methods listed here. The whole Section 4 is devoted to the discussion of TBS.

The above approaches only assumes partial knowledge of the graph. When we have full knowledge of a graph, sampling with contraction may be another approach. It can be used to support visualization or accelerate followup algorithms. The following two types are less discussed in previous literature but we think they are promising in this context:

- Edge Sampling with Contraction (ESC). This is an iterative process. Each round one edge $(u,v) \in E$ is sampled and let
    - $V \leftarrow V + \{(u,v)\} - \{v\} - \{u\}$
    - $\delta((u,v)) = \{((u,v),w) | (u,w) \in \delta(u) \cup (v,w) \in \delta(v)\}$
    - $E \leftarrow E + \delta((u,v)) - \delta(u) - \delta(v)$.

- Vertex Sampling with Contraction (VSC). Similar to ESC, but one vertex $v \in V$ is sampled each round and vertices in $\{v\} \cup N(v)$ are contracted into one vertex with corresponding edges. This is equivalent to perform edge contraction on all $e \in \delta(v)$. One will find that VSC is a more constrained version of ESC which introduce dependence between contracted edges.

### 3.3.2 Relations between Sampling Approaches

VS and ES are simple and suitable for theoretical analysis. This is because samples taken from VS and ES are uncorrelated. Independence makes it possible for us to apply concentration bounds, e.g. Chernoff bound (Appendix A). However, in most real applications, one can not perform VS and ES directly due to all kinds of constraints, e.g. can not enumerate the ID space. In this case, TBS becomes more practical, which only relies on a small sized starting topology (e.g. a few seed nodes) and expand it during the exploration. Note that VS, ES and TBS are not totally different from each other. Certain TBS techniques can be used to generate VS or ES:

- Random Walk (RW) (Section 4.3) results in uniform edge distribution on an undirected graph.
- Metropolis-Hastings Random Walk (MHRW) (Section 4.4) results in uniform vertex distribution. In fact, the Metropolis-Hastings algorithm [42] can tailor the Markov chain to any vertex distribution.

These two observations are the bridges between theoretical analysis and practical implementation. The general methodology is:

- Analyze the properties with simple VS, ES, or its variants.
- Construct equivalent VS or ES samplers using TBS techniques.

Here are a few remarks for this methodology:

- If TBS is used to mimic VS and ES, we usually spend substantially more probes. For example, starting from an arbitrary node, we need to run random walk for the mixing time in order to reach the stationary distribution. This is also known as "burn in" period of the Markov chain. After reaching the stationary distribution, researchers often take samples from the random process with certain gap, which results in a waste of resources.
- Although samples taken from TBS have correlations, this can sometimes be utilized to construct more efficient estimators for certain properties, e.g. [20] for clustering coefficient. Approaching from TBS directly can reduce the waste in mimicking VS and ES. This makes TBS based property estimator of its own interest.

## 4 Traversal Based Sampling

Traversal Based Sampling has a very long history and is still the research focus in recent years. It is also called "topology based sampling" [3] or "sampling by exploration" [36]. In this section, we present well studied TBS techniques from the classical ones to recent ones. Some less popular or too recent methods/ improvements are briefly mentioned in Section C.

Traversal Based Sampling has a very long history and is still the research focus in recent years. It is also called "topology based sampling" [3] or "sampling by exploration" [36]. In this section, we present well studied TBS techniques from the classical ones to recent ones.

### 4.1 Breadth/ Depth/ Random First Sampling (B-/D-/R- FS)

One recent paper [11] formalized the framework for three intuitive graph traversal methods. Since Breadth First Sampling and Depth First Sampling are deterministic, we also use their classical name Breadth First Search and Depth First Search, respectively. The framework works as follows:

- Initialize the queue with a starting vertex: $Q \leftarrow \{v_0\}, V_s \leftarrow \{v_0\}$. Let $L = \{\}$ be the set of visited vertices.

- Loop until budget $B$ is exhausted:
    - Dequeue one element: $v = Q.\text{dequeue}()$
    - $B \leftarrow B - b, L \leftarrow L + \{v\}$
    - For $u \in N(v) \cap u \notin L \cap u \notin Q$, invoke $Q.\text{enqueue}(u)$

The sampled vertices are $V_s = L$ and the edges are dependent on the context. The difference between BFS, DFS and RFS is in the implementation of "dequeue()" methods. For BFS, the first element is selected; For DFS, the last element is selected; For RFS, a random element is selected.

## 4.2 Snow-Ball Sampling (SBS)

Snowball Sampling [17] has long been used in sociology studies, where an investigation is performed on a hidden population (e.g. drug abusers). It is also called "network sampling" [12][57] or "chain referral sampling" in many literatures (as counterparts of "straight sampling", "stratified sampling", etc). A stage $t$ name $k$ SS is defined as:

- Start from initial set of vertices $V^{(0)}$. This set may be obtained through a random sample on vertices or side knowledge of the hidden population.

- At stage $i$, ask $\forall v \in V^{(i-1)}$ to name $k$ neighbouring nodes. How to name varies according to the specific study. This is equivalent to obtain a sample of incident edges of $V^{(i-1)}$. We denote those edges by $E^{(i)}$. Vertices observed in this stage are $\tilde{V}^{(i)} = \{u, v | (u, v) \in E^{(i)}\}$. We put new vertices into stage $i$, i.e. $V^{(i)} = \tilde{V}^{(i)} - \bigcup_{j=0}^{i-1} V^{(j)}$.

- The process continues for $t$ stages. The sampled graph is $G_s = <V_s, E_s>$ where $V_s = \bigcup_{j=0}^{t} V^{(j)}$ and $E_s = \bigcup_{j=1}^{t} E^{(j)}$.

Note that we output the whole sampled graph to make our discussion unified. In sociology study, people only care $V_s$ for most of the time. As long as $V_s$ is a representative set of vertices of certain population, statistics can be done on this group of people, e.g. finding the average opinion on some issues.

One remark is that SBS is very similar to BFS. BFS exhaustively expand the neighbourhood of current vertex while SBS only expands a fixed number of them.

## 4.3 Random Walk (RW)

Random Walk also arises from different context. The general description is:

- Start from an initial vertex $v^{(0)}$.

- At step $i$, choose one neighbouring vertex $u \in N(v^{(i-1)})$ (maybe uniformly at random or according to some weight). Let $v^{(i)} \leftarrow u$ be the next vertex and include this edge $\tilde{E}_s \leftarrow \tilde{E}_s + \{(v^{(i-1)}, v^{(i)})\}$.

- Repeat $t = \frac{B}{b}$ steps. Return the sampled graph $G_s = <V_s, E_s>$. There are two possibilities of the vertex set and edge set depending on different application:
    - $V_s = \{v^{(0)}, v^{(1)}, ... v^{(t)}\}$ and $E_s = \tilde{E}_s$. This corresponds to the scenario that neighbourhood of vertices are unknown, but we can somehow probe one neighbour according to some rules.
    - Denote $\tilde{V}_s = \{v^{(0)}, v^{(1)}, ... v^{(t)}\}$. Then $E_s = \bigcup_{v \in \tilde{V}_s} \delta(v)$, and $V_s = \{u, v | (u, v) \in E_s\}$. This is just the VSN technique corresponding to many network crawling scenarios. We simply treat RW as one way to perform VS and obtain $\tilde{V}_s$.

RW is related with SBS in the following aspects:

- One can regard RW as performing SBS with $k = 1$ and the "naming" criterion for participants is "pick $k = 1$ of your random neighbour". **However**,

- RW is memoryless. In SBS, participants from previous stages are excluded. In RW, one can revisit some vertices. The memoryless property of RW makes it more appealing for theoretical analysis, e.g. obtain stationary distribution from Markov Chain analysis. It turns out that many analyses done on SBS are actually done on RW, especially in the discussion of RDS (Section 4.9).

We have a few remarks for RW:

- It can be shown [40] that RW on undirected graph (connected and non-bipartite) results in uniform distribution on edges. Towards this end, uniform ES can be made practical via RW in many scenarios.
- A vertex has degree-proportional probability to be in $V_s$, i.e. $\pi(v) = \frac{d(v)}{m}$, but the transition probability between vertices can be varied to output uniform VS. This is the problem that Metropolis-Hastings Random Walk (Section 4.4) want to address.
- It's shown [52] (for degree distribution) that the error of estimators is inverse proportional to the spectral gap of RW transition matrix $D^{-1}A$. This result is intuitive because the spectral gap encodes the conductance of the graph. The larger the conductance, the less likely a random walker gets stuck in a local region. For graphs with small spectral gap, MIRW (Section 4.6) and MDRW (Section 4.7) are proposed.

## 4.4 Metropolis-Hastings Random Walk (MHRW)

Metropolis-Hastings algorithm [42] was widely used in Markov Monte Carlo Chain to obtain a desired vertex distribution from an arbitrary undirected connected graph. Denote the transition probability from $x$ to $y$ by $P_{x,y}$ and the desired vertex distribution by $\pi_x$. The most general form to our knowledge is: ([43]*)

$$P_{x,y} = \begin{cases} M_{x,y} \times \min\left\{1, \frac{\pi_y}{\pi_x}\right\} & x \neq y \cap y \in N(x) \\ 0 & x \neq y \cap y \notin N(x) \\ 1 - \sum_{y \neq x} P_{x,y} & x = y \end{cases}$$

where $M_{x,y} = M_{y,x}$ is a normalization constant for the pair of $x$, $y$ satisfying $\sum_{y \neq x} P_{x,y} \leqslant 1$. When we let $M_{x,y} = M \leqslant \min_{v \in V} \frac{1}{d(v)}$, this degrades to the result given by [43]. Note that adding more (higher weight) self-loops will make the mixing time longer, so it is better to choose a larger $M_{x,y}$ when possible. One choice is $M_{x,y} = \min\left\{\frac{1}{d(x)}, \frac{1}{d(y)}\right\}$. If we further let $\pi_x = \pi_y = \frac{1}{n}$, Eq. 4.4 degrades to the result quoted in [25][50][15].

To prove the correctness, one can first observe it is a valid conditional probability distribution and then show that $\pi_x P_{x,y} = \pi_y P_{y,x}$. Since $\pi_x$ sums up to 1, the time reversibility is satisfied. That means $\pi_x$ is indeed the stationary distribution of this Markov Chain.

One remark is that the application scenario of MHRW is more limited than ordinary RW. In order to calculate the transition probability one has to know the degree of neighbouring vertices. This information is in general unavailable. Nevertheless, there are some scenarios where this is possible. For example, in a P2P network, the number of peers is usually a fixed parameter. For another example, the API of some OSNs may return friend list with rich information besides mere IDs (e.g. number of followers and followees).

## 4.5 Random Walk with Escaping (RWE)

There is one widely used variation of the above mentioned RW. Besides walking to neighbours, the random walker can jump to an arbitrary random node $u \in V$. In original Pagerank [9] computation, this is to 1) make the chain both aperiodic and irreducible; 2) enlarge the eigen gap in order to converge faster. This technique is termed as "random jump" in [36][25]. However, RWE is not a very meaningful technique in graph sampling. TBS in general solves the problem where the whole graph can not be reached, or at least direct VS or ES is hard. RWE assumes the existence of efficient VS, which renders it less usable in many scenarios.

As a supplementary note, RWE's stationary distribution is obtained by solving the equation

$$\pi = \alpha W \pi + (1-\alpha)\frac{1}{n}\mathbf{1}$$

where $W = AD^{-1}$ gives the transition probability to neighbours. It gives

$$\pi = \frac{1}{n}(1-\alpha)[I - \alpha W]^{-1}\mathbf{1}$$

The inversion makes $\pi_v$ correlated with other entries in $W$, which is unavailable unless we have the full graph. So even we can run RWE on the graph, it is hard to construct unbiased estimators for graph properties because we do not know $\pi$ (Section 4.9 gives the way to correct bias if $\pi$ is unknown). The only simple conclusions we can get lie on the two ends of $\alpha$'s domain: 1) when $\alpha = 0$, $\pi$ is uniform; 2) when $\alpha = 1$, $\pi$ is degree proportional.

In [25], the authors proposed the "Albatross Sampling" algorithm. This is essentially RWE presented above. The only difference is that $W$ is constructed by MHRW instead of simply $W = AD^{-1}$.

## 4.6 Multiple Independent Random Walkers (MIRW)

One problem of RW is that it is prone to be trapped into local dense region. Thus it is believed to have high bias according to different initial vertex. Note that this phenomenon is welcome in some applications like local graph partitioning ([33] L10), Sybil detection [69], etc. However, in many other applications, especially graph property estimation via random walks, we want to alleviate the problem.

MIRW [15] was proposed to address this problem. Later, [51] shows that MIRW actually results in higher estimation errors and proposes MDRW (Section 4.7). We document MIRW here for completeness. First, we perform VS to get $l$ initial vertices. Then we split the budget to $l$ random walkers and let them execute independently. To make it simple, let all random walkers to walk $t = \frac{B}{lc}$ steps. In stationary state $(v_1, v_2, ...v_l)$, the joint distribution on vertices is

$$P(v_1, v_2, ...v_l) = \prod_{i=1}^{l} \frac{d(v_i)}{m}$$

and the joint distribution on edges is

$$P(e_1, e_2, ...e_l) = \prod_{i=1}^{l} \sum_{v:e \in \delta(v)} \Pr\{e_i|v\}\Pr\{v\} = \prod_{i=1}^{l} \frac{1}{m} = \left(\frac{1}{m}\right)^l$$

The problem for MIRW is that $P(v_1, v_2, ...v_l)$ can deviate from the uniform distribution $\tilde{P}(v_1, v_2, ...v_l) = \frac{1}{n^l}$ by a considerable large amount. The difference is significant when vertex degree span a large range, which is the case for many real social networks. One can not use it to estimate vertex labels.

## 4.7 Multi-Dimensional Random Walk (MDRW)

MDRW [51] is also called Frontier Sampling (FS). An $l$-dimensional RW works as follows:

- Initialize $L = (v_1, v_2..., v_l)$ with $m$ random vertices (e.g. via VS).
- At each step, choose one vertex $v_i \in L$ with the probability $p(v_i) \propto d(v)$. Choose a random neighbour $u \in N(v_i)$. Assign $E_s \leftarrow E_s + (v_i, u)$ and $L \leftarrow (v_1, ...v_{i-1}, u, v_{i+1}, ..., v_l)$.
- Loop until budget is used out. Output $E_s$ as the sampled edges.

We define the $l$-the Cartesian power of $G$, $G^l = <V^l, E_l>$, as follows:

- $V^l = \{(v_1, v_2, ...v_l)|v_1, v_2, ...v_l \in V\}$
- $E_l = \left\{(u, v)|u = (u_1, u_2, ...u_l) \in V^l, v = (v_1, v_2, ...v_l) \in V^l, \exists i \, s.t. \, (u_i, v_i) \in E, \bigcap_{j \neq i} u_j = v_j\right\}$

The $l$-dimensional RW on $G$ is equivalent to a 1-dimensional RW on $G^l$. [51] shows that MDRW yields better estimators some graph properties. [41] says that MDRW yields uniform distribution on both edges and vertices when $l \to \infty$.

## 4.8 Forest Fire Sampling (FFS)

Forest Fire was originally proposed in [37] as a graph generation model which captures some important observations in real social networks, like densification law, shrinking diameter, and community guided attachement. In [36], the author adapted this graph generation model to perform graph sampling and termed it as Forest Fire Sampling (FFS). It was later abbreviated as "forest fire" in many followup graph sampling works.

FFS is a probabilistic version of SBS (see Section 4.2). In SBS, $k$ neighbours are selected at each round. In FFS $K \sim \text{Geometric}(p)^1$ number of neighbours are selected. SBS and FFS are linked by setting $p = \frac{1}{k}$. Then we have $\mathbb{E}[K] = k$.

One remark is that apart from the above analysis, FFS is still more close to SBS than other RW variations. In RW and its variations, repetitions are included in the sample for estimation purpose. When FFS is used to do estimation, the algorithm avoids previously selected vertices (once a vertex is "burned", it will no longer be burned again). This setting is same as original SBS where nodes selected in previous stages are excluded from current stage.

## 4.9 Respondent Driven Sampling (RDS) (RWRW)

RDS was first presented in sociology studies to perform estimation on hidden population, e.g. [22][54]. This technique becomes hot recently and sees applications in many other fields. The original core idea is to run SBS and correct the bias according to the sampling probability of each vertex in $V_s$. Many people now implement SBS as RW because the stationary distribution derived from RW can be easily used to correct the bias. For this reason, [15] also termed this technique as Re-Weighted Random Walk (RWRW).

Note that RDS (RWRW) and MHRW are often compared in recent literatures, e.g. [15][50], leaving the confusion that RDS is a sampling technique. Indeed RDS itself is not a standalone sampling or say graph exploration technique. It either uses SBS (original operation) or RW (current trend) for sampling. After that, it proposes to use Hansen-Hurwitz estimator [19] to **correct the bias of an estimator for the mean of vertex label distribution** (Section 5.1). It does not matter how we get the samples (by VS, ES, or TBS). As long as we know the sample probability, we can invoke the bias correction technique. Since MHRW can generate a uniform sample for $V$, direct estimator can be applied on the sampled vertices. By looking at the sampling+estimating procedures as a whole, RWRW and MHRW seem to have the same objective and similar results. [16] made this relation more clear.

Although RDS is not a standalone TBS technique in our terminology, we also note the idea here for easier reference. Suppose we have a sample $|V_s| = n_s$ drawn with the distribution $\pi(v)$, $\forall v \in V$. Repeated samples of a single vertex is allowed. Now we want to estimate the parameter $\theta = \frac{1}{n} \sum_{v \in V} g(v)$, where $g : V \to \mathbb{R}$ is a function to generate vertex labels (e.g. degree, $g(v) = d(v)$). Note that the naive estimator $T_1 = \frac{1}{n_s} \sum_{v \in V_s} g(v)$ is not consistent. As $n_s \to \infty$, $T_1 \to \sum_{v \in V} \pi(v) g(v)$. Now we want to substitute $g(v)$ with $h(v)$ in the expression of $T_1$ with the expectation that

$$n_s \to \infty \Rightarrow T_2 = \frac{1}{n_s} \sum_{v \in V_s} h(v) \to \theta = \frac{1}{n} \sum_{v \in V} g(v)$$

One can show that one choice of $h(v)$ is given by

$$h(v) = \frac{g(v)}{n \, \pi(v)}$$

---

1. In original paper [37], it was said to be a binomially distributed value with certain expectation. A binomial distribution is characterized by two parameters $n$, $p$ but only one parameter was given here. In [3], the authors interpreted it as a geometric distributed r.v. and the given expectation is enough to characterize it. We adopt the interpretation by [3].

When $n_s \to \infty$

$$T_2 = \frac{1}{n_s}\sum_{v \in V_s} h(v) = \mathrm{E}_\pi\left[\frac{g(v)}{n\,\pi(v)}\right] = \sum_{v \in V} \frac{g(v)}{n\,\pi(v)}\pi(v) = \frac{1}{n}\sum_{v \in V} g(v)$$

This estimator is still too general to use. There are two difficulties:

- $n$ may be unknown in many scenarios. We must estimate $n$ using $\pi$. By setting $g(v) = 1$, $\forall v \in V$, we know $\frac{1}{n_s}\sum_{v \in V_s} \frac{1}{n\,\pi(v)}$ is a consistent estimator for 1, namely,

$$\frac{1}{n_s}\sum_{v \in V_s} \frac{1}{n\,\pi(v)} = 1 \Rightarrow \frac{1}{n_s}\sum_{v \in V_s} \frac{1}{\pi(v)} = n$$

in probability. So we find the consistent estimator for $n$

$$\hat{n} = \frac{1}{n_s}\sum_{v \in V_s} \frac{1}{\pi(v)}$$

and let our estimator for parameter $\theta$ be

$$\hat{\theta} = \frac{1}{n_s}\sum_{v \in V_s} \frac{g(v)}{\hat{n}\,\pi(v)}$$

- The stationary distribution output by our RW is $\pi(v) = \frac{d(v)}{m}$. $d(v)$ can be obtained during the TBS procedure someitmes. However, $m$ is unknown in most scenarios. Luckily, with the above estimator $\hat{n}$, $m$ just cancels out and we get:

$$\hat{\theta} = \frac{1}{n_s}\sum_{v \in V_s} \frac{g(v)}{\frac{d(v)}{m}}\frac{1}{\frac{1}{n_s}\sum_{v \in V_s} \frac{1}{\frac{d(v)}{m}}} = \frac{1}{\sum_{v \in V_s}\frac{1}{d(v)}}\sum_{v \in V_s} \frac{g(v)}{d(v)} \qquad (2)$$

After addressing the two problems, RDS combined with RW is a practical approach to estimate graph properties without knowing the full graph. By substituting $g(v)$ with application specific functions, we get the formula used in [16][15][50].

One remark is that the correction method presented by RDS is not tied to RW. As long as we can draw samples from a distribution $\pi$ and know the relative ratio between all $\pi(v)$ and $\pi(u)$ for $v, u \in V_s$, similar estimator can be constructed.

## 5 Graph Properties

Given a graph $G$, a property is defined as a function $f(G)$, where $f$ is probably a vector function in many cases. Note that this definition is different from the classical discussion of random graphs (e.g. [10][33, L5] ), where it is defined as a subset of graphs from the family, i.e. $Q \subseteq \mathcal{G}$. If $G$ is an r.v., so is $f(G)$. The classical definition is just an indicator for a certain $f(G)$. For example, define $Q$ as "subset of graphs who are connected"; Define $f(G): G \to \#$ of connected components; Obviously $G \in Q \Leftrightarrow \mathrm{I}[f(G) = 1]$. Towards this end, we'll use the former definition, which is more general and implicitly adopted in a wide spectrum of works.

### 5.1 Vertex/ Edge Label Distribution

Before we discuss concrete graph properties, we first present two general type of properties: vertex label and edge label [51]. Suppose there is a collection of labels $\mathcal{L}$. Every vertex $v$ is associated with a set of labels $L(v) \subseteq \mathcal{L}$. The label distribution on $\mathcal{L}$ is defined as

$$p(l) = \frac{\sum_{v \in V} \mathrm{I}[l \in L(v)]}{\sum_{v \in V} |L(v)|}$$

where I[·] is the indicator function. For example, let $\mathcal{L}$ be all non-negative integers and let $L(v) = \{d(v)\}$. Then $p(l)$ becomes the degree distribution. Actually, some of the properties discussed in Section 5.2 are just vertex label distribution.

The edge label distribution can be defined similarly but it is seldom seen. The reason may be that most meaningful properties derived from mere topology are defined over vertices, e.g. degree distribution and clustering coefficient. Nevertheless, there are some application specific properties which can be viewed as edge label distribution. For example, [50] uses two TBS techniques, RDS and MHRW (see Section 4), to sample a P2P overlay. One of the properties the authors measure is Round Trip Time (RTT) between peers. This property is defined over edges and can be obtain at every peer query.

Here are a few remarks to vertex/ edge label distribution:

- Label distribution gives a general framework to define a class of properties. This definition allows a straightforward construction of estimators. For example, a graph property $f(G)$ can be calculated as $f(V)$ if it is a vertex label distribution. Naturally, $f(V_s)$ gives an estimator of this property if $V_s$ is a uniform sample from $V$. Edge label distribution is similar. In other words, how well a sampler preserves the property becomes how well the sampler generates a uniform distribution over $V_s$ or $E_s$.

- The above statement assumes that labels can be efficiently acquired as long as we sample a vertex. This is true for some cases, e.g. age, gender and other attributes crawled from profiles of OSN. However, many other properties are not readily available upon each probing. The information acquired also differs across sampling algorithms. What's more, some properties when associated with the type of the graph can yield special results. Towards this end, a discussion of concrete properties is still needed even if they are covered by label distribution.

## 5.2 Classical Graph Properties

In order to characterize a graph, people have proposed dozens of widely used graph properties. Graphs share the same properties are believed to be "similar". Indeed, by observing or estimating those properties, people can already say something about the graph. e.g. [6] performed classification on synthesized networks using 40+ features (variants of some of the following properties) and found that some types of networks are well separable from others. In this section, we collect most of the text-book like properties.

- Network Size. Two simple values to describe network size are number of vertices $n = |V|$ and number of edges $m = |E|$.

- Degree Distribution (Deg). Randomly pick a node $X \in V$, let

$$p_{\text{deg}}(k) = \Pr\{d(X) = k\}$$

  $p_{\text{deg}}(k)$ is thus the p.d.f. for the degree distribution.

- Average Degree (AD). Average degree is the expectation

$$E[d(X)] = \sum_k k\, p_{\text{deg}}(k)$$

- Power Law Exponent (PLE). After obtaining the degree distribution, one can fit the power law exponent $\gamma$, s.t. the fitted distribution $p_{\text{fit}}(k) \propto k^{-\gamma}$ is most close to observed degree distribution $p_{\text{deg}}(k)$. How to fit is beyond the scope of this paper, see [49] for more info. This property only makes sense when the original and sampled graph is close to power-law graphs.

- Graph Density (GD). It is defined as the ratio of observed number of edges over maximum possible number of edges:

$$\text{density} = \frac{m}{\binom{n}{2}} = \frac{2m}{n(n-1)}$$

Since $2m = \sum_k p_{\deg}(k) \times n \times k = E[d(X)] \times n$, GD is just a summary metric of Deg.

- Path Matrix (PM). Denote the adjacency matrix by $A$, the $t$ step path matrix is:

$$[P_t]_{i,j} = \begin{cases} 1 & \exists_{\tau \leqslant t} [A^\tau]_{i,j} > 0 \\ 0 & \text{else} \end{cases}$$

That is the path matrix encodes the reachability information between all pairs of vertices.

- Shortest Path Matrix (SPM). The shortest path matrix can be defined on PM as

$$[S]_{i,j} = \arg\min_t \{[P_t]_{i,j} = 1\}$$

namely, $S$ records all pair shortest path distance.

- Average Path Length (APL). APL refers to the average shortest path length:

$$\text{APL} = \frac{1}{\binom{n}{2}} \sum_{1 \leqslant i < j \leqslant n} S_{i,j} = \frac{2}{n(n-1)} \sum_{1 \leqslant i < j \leqslant n} S_{i,j}$$

- Closeness Centrality (CloC) [66]. The farness of a vertex is defined as the sum of distances to all other vertices. The closeness centrality is just its inverse:

$$\text{Closeness}(v) = 1 / \sum_{u \in V} S_{v,u}$$

- Radius (R) [38]. Radius of a vertex is the maximum shortest path distance to all other vertices. We can define it using SPM:

$$R(v) = \max_{u \in V} S_{v,u}$$

- Diameter (Dia) [38]. Diameter is defined as the maximum radius of all vertices, i.e. $D(G) = \max_{v \in V} R(v)$. In other words, it measures the longest distance between all pair of vertices. In practice, the network may be disconnected or have a few large radius vertices. Diameter may not be a meaningful property. Instead, effective diameter is used, i.e. the distance within which $\alpha$ fraction (e.g. $\alpha = 90\%$) of the pairs can reach each other.

- Betweenness Centrality (BC) [66] is the number of shortest paths a vertex is involved in. Denote the collection of shortest paths between $s$ and $t$ by $\sigma_{s,t}$. The paths $v$ is involved in is $\sigma_{s,t}(v) = \{p \in \sigma_{s,t} | v \in p\}$. Then the betweenness centrality is defined as:

$$\text{Betweenness}(v) = \sum_{s \neq v, t \neq v} \frac{|\sigma_{s,t}(v)|}{|\sigma_{s,t}|}$$

- Assortativity (As). Assortativity is generally defined as the Pearson Correlation of similarity between neighbouring vertices. One can use degree as the similarity measure [48] and leads to the following definition. The distribution of remaining edges (except for the one that link the two vertices under consideration) is: $q_k = \frac{(k+1)p_{\deg}(k+1)}{\sum_j j p_j}$. Define the joint distribution of the degree of two vertices by $e_{j,k}$. Then the assortativity is defined as:

$$r = \frac{1}{\sigma_q^2} \sum_{j,k} jk(e_{j,k} - q_j q_k)$$

- Clustering Coefficient (CC). For the node $v$, the clustering coefficient is defined as

$$C(v) = |\Delta(v)| / \binom{|N(v)|}{2}$$

where $\Delta(v) = \{(u,w) | u \in N(v), w \in N(v), (u,w) \in E\}$ is the set of observed edges in neighbourhood. $\binom{|N(v)|}{2}$ is the number of all possible edges. One can also interpret it as the ratio of observed triangles over all possible triangles on $v$'s ego-network.

- Average Clustering Coefficient (ACC). The average clustering coefficient (or Global Clustering Coefficient [51]) is a summary property of the CC of all vertices:

$$C(G) = \frac{1}{n} \sum_{v \in V} C(v)$$

- Global Clustering Coefficient (GCC). GCC is defined as [20]:

$$\mathrm{GCC}(G) = \frac{|\{<u,v,w>|(u,v),(v,w),(w,u) \in E\}|/3}{|\{<u,v,w>|(u,v),(v,w) \in E\}|/2}$$

It is also called transitivity [55]. Note that there may be constant differences in the definition across literatures.

- Self Similarity (SS). [39] proposed a probably more precise definition to capture "scale-free" property of some networks. Let $s(G) = \sum_{(u,v) \in E} d(u)d(v)$. This value is higher is vertices are connected to similar other degree vertices. Denote $s_{\max} = \max_{G \in \mathcal{G}} s(G)$, where $\mathcal{G}$ is the family of graphs that have same degree distribution as $G$. The self-similarity of $G$ is defined as: (digested from [67])

$$S(G) = \frac{s(G)}{s_{\max}}$$

This value is positively related with the assortativity defined above.

- Spectrum (S). The spectrum of a matrix is the eigenvalue distribution. For a graph, several matrices may be used [23]: Adjacency matrix $A$; Laplacian $L = D - A$; Normalized adjacency matrix $\mathcal{A} = D^{-1/2} A D^{-1/2}$; Normalized Laplacian matrix $\mathcal{L} = I - \mathcal{A}$; Left normalized adjacency matrix $A_{\text{left}} = D^{-1} A$; Left normalized Laplacian matrix $L_{\text{left}} = I - A_{\text{left}}$. They have different physical meanings and different application scenario. The spectrum encodes certain graph properties, e.g. the eigen gap of $\mathcal{L}$ encodes the conductance of the graph.

- Largest Eigen Vector (LEV). The largest eigen vector of $A$, $\mathcal{A}$ and $A_{\text{left}}$ encodes a type of centrality information of vertices. For example, left eigen vector of $A_{\text{left}}$ is just the PageRank value [9]. It is also the eigenvector of $\mathcal{A}$ scaled by $D^{1/2}$. The eigenvector of $A$ is defined as the eigenvector centrality [47].

- Smallest Eigen Vector (SEV). The smallest **non-zero** eigen vector of $L$, $\mathcal{L}$, and $L_{\text{left}}$ encodes a type of partition information. For example, 2nd eigen vector of $\mathcal{L}$ will be piecewise linear if the graph has several connected components [62] (ideal partition). We address "non-zero" because the smallest eigen value is 0 and corresponding all one's vector is not interesting ($L\mathbf{1} = \mathbf{0}$).

## 5.3 Classical Properties Studied in Literatures

It turns out that many of the properties were numerically studied after certain sampling methods. However, only a few properties are theoretically studied. In the later part of this chapter, we present two theoretical results we get during the survey. We summarize those properties in the surveyed works in the following Table 1. It may provide some pointers to potential future works. [2]

---

2. We can not afford a thorough scan of all properties. There are many properties appear in research works but are not so widely used. Interested readers can consult those literatures. e.g. [36] presented some dynamic properties; [6] presented 10 other summary properties which are not included in this section; See [55] for generalized clustering coefficient; [64] is a classical textbook on social network analysis, including the discussion of many classical properties.

| Property | Numerical Study | Theoretical Study |
|---|---|---|
| NS | [28][20] | [28][20] |
| Deg | [60][59][35][51][15][25][27][63][6] [36][30][3][4][31][52][34][68] | [60][59][35][51][24][30][31][52][34] |
| AD | [11][28][4][31] | [24][28] |
| PLE | [11] | |
| GD | [11][28] | [28] |
| R | | |
| Dia | [11][6][36] | |
| PM | | |
| SPM | [6][3][4] | |
| CloC | [6] | |
| APL | [35][6][3][4] | |
| BC | [35][6] | |
| As | [35][11][51][15] | [51] |
| CC | [15][6][36][3][4] | |
| ACC | [35][51][15][63][6][36][4][20] | [51][20] |
| GCC | [20] | [20] |
| SS | | |
| S | [36] | |
| LEV | [27][6][36] | |
| SEV | | |

**Table 1.** Studied Properties in Past Literatures

We can not afford a thorough scan of all properties. There are many properties which appeared in research works but are not so widely used. Interested readers can consult those literatures. e.g. [36] presented some dynamic properties; [6] presented 10 other summary properties which are not included in Section 5.2; See [55] for generalized clustering coefficient; [64] is a classical textbook on social network analysis and it includes the discussion of many classical properties.

### 5.4 Advanced Graph Properties

Classical properties are well established in past years. They are useful for understanding the graphs. However, those graph are less useful for further supporting graph algorithms. In this section, we discuss some advanced graph properties. Those properties are more rich in terms of capturing the structure of graphs and they have been used as objectives of graph algorithms. By preserving those properties on a sampled graph, we can systematically accelerate those algorithms.

#### 5.4.1 Cut and Ratio-/ Normalized-/ Weighted- Cut

The cut of a set $S$ is defined as the edges crossing $S$ and $V - S$:

$$\text{Cut}(S) = |\delta(S)| = |\{(u,v) \in E | u \in S, v \notin S\}|$$

The ratio cut takes the size of $S$ into consideration:

$$\text{RCut}(S) = \frac{|\delta(S)|}{|S|}$$

The normalized cut further considers the importance of vertices in terms of degree:

$$\text{NCut}(S) = \frac{|\delta(S)|}{|\text{vol}(S)|}$$

A most general quantity of this category is the weighted cut:

$$\text{WCut}(S) = \frac{|\delta(S)|}{\sum_{v \in S} w(v)}$$

where $w(v) \in \mathbb{R}, \forall v \in V$ is the weighting function. If $w(v) = 1, \forall v \in V$, this degrades to RCut. If $w(v) = d(v), \forall v \in V$, this degrades to NCut.

The cut series can very well capture the structure of the graph. For example, when $S$ is a singleton $\{v\}$, the cut degrades to $N(v)$. The cut distribution thus subsumes the degree distribution. Cut series arise as building blocks of optimization objective for a lot problems (e.g. NCut [56] for image segmentation). Interested readers can refer to [23] for the application on spectral clustering.

### 5.4.2 Association and Ratio-/ Normalized-/ Weighted- Association

Association is the counterpart of cut. It is defined as the edges within one set of vertices. The other variants can also be defined accordingly:

$$\begin{aligned}
\text{Assoc}(S) &= |\text{vol}(S) - \delta(S)| \\
\text{RAssoc}(S) &= \frac{\text{Assoc}(S)}{|S|} \\
\text{NAssoc}(S) &= \frac{\text{Assoc}(S)}{|\text{vol}(S)|} \\
\text{WAssoc}(S) &= \frac{\text{Assoc}(S)}{\sum_{v \in S} w(v)}
\end{aligned}$$

See the reference in [23] for some applications using those quantities. To facilitate further discussion, we also denote $\rho(S) = \text{vol}(S) - \delta(S)$.

### 5.4.3 Conductance and Expansion

Conductance and expansion are similar quantities to NCut and RCut. They are defined as:

$$\begin{aligned}
\text{Conductance}(S) &= \frac{\delta(S)}{\min\{\text{vol}(S), \text{vol}(V - S)\}} \\
\text{Expansion}(S) &= \frac{\delta(S)}{\min\{|S|, |V - S|\}}
\end{aligned}$$

Due to the min operator, they are less used as optimization objective. In some discussions, people restrict the set $S$ to be $|S| \leqslant \frac{|V|}{2}$ or $|S| \leqslant \frac{|\text{vol}(V)|}{2}$. In this way, the min operator vanishes and they are same as RCut and NCut.

### 5.4.4 Quadratic Form

The quadratic form of adjacency matrix $A$ and Laplacian $L = D - A$ are more general quantity to characterize the graph. Denote the characteristic vector for a set of vertices $S$ by:

$$\chi_S(v) = \begin{cases} 1 & v \in V \\ 0 & v \notin V \end{cases}$$

One can see that $\chi_S^\mathrm{T} A \chi_S$ encodes the association of set $S$ and $\chi_S^\mathrm{T} L \chi_S$ encodes the cut of set $S$. Towards this end, the quadratic form is a more general notion than the cut series and association series.

By using different matrix and weighted characteristic vector, we can encode other variants of cut and association. For example, one can expand the quadratic form of Laplacian as

$$x^\mathrm{T} L x = \sum_{(u,v) \in E} (x(u) - x(v))^2$$

We can set the characteristic vector to be

$$\chi_S(v) = \begin{cases} \dfrac{1}{\sqrt{|S|}} & v \in S \\ 0 & v \notin S \end{cases}$$

and then

$$\chi_S^\mathrm{T} L \chi_S = \sum_{(u,v)\in\delta(S)} \left(\frac{1}{\sqrt{|S|}} - 0\right)^2 = \frac{|\delta(S)|}{|S|}$$

This is just the ratio cut. Other quantities can be obtained similarly.

In graph sparsification problems, a very widely used criterion for $\varepsilon$-approximation is: ([58] ch6.4)

$$(1-\epsilon)x^\mathrm{T} L_G x \leq x^\mathrm{T} L_H x \leq (1+\epsilon) x^\mathrm{T} L_G x, \forall x \in \mathbb{R}^n$$

According to the analysis in this section, we also propose to use adjacency matrix as one alternative criterion: (*)

$$(1-\epsilon)x^\mathrm{T} A_G x \leq x^\mathrm{T} A_H x \leq (1+\epsilon) x^\mathrm{T} A_G x, \forall x \in \mathbb{R}^n$$

Adjacency matrix encodes yet another type of information (e.g. association) which is not readily available from Laplacian. Thinking along this line [3], the quadratic of degree matrix is also useful: (*)

$$(1-\epsilon)x^\mathrm{T} D_G x \leq x^\mathrm{T} D_H x \leq (1+\epsilon) x^\mathrm{T} D_G x, \forall x \in \mathbb{R}^n$$

It encodes the volume of a set of vertices. If a sampling procedure can preserve those quadratic forms, a larger class of algorithms may apply.

### 5.4.5 Modularity

Modularity is a classical quality measure in community detection problems [46]. We use the form presented in [1]. Let $\mathcal{C} = \{C_1, C_2, ... C_n\}$ be a clustering of the graph s.t. $C_i \cap C_j = \emptyset$ and $C_1 \cup C_2 ... \cup C_n = V$. The modularity of the whole graph can be expressed as:

$$Q(\mathcal{C}) = \frac{1}{2m} \sum_{C \in \mathcal{C}} \sum_{u \in C, v \in C} \left(A_{u,v} - \frac{d_u d_v}{2m}\right)$$

The term $\frac{d_u d_v}{2m}$ computes the expected number of edges between $u, v$ on a Fixed Degree Distribution Random Graph (FDDRG). Thus, for each cluster, it computes the deviation of observed graph from a random graph. If a set of vertices have closer relationships, their Modularity should be higher. This fact makes Modularity a widely accepted metric for community detection problems.

To make it a property similar to cut series and association series, we can define the Modularity for a set of vertices: (*)

$$\mathrm{Modularity}(S) = \sum_{u \in S, v \in S} \left(A_{u,v} - \frac{d_u d_v}{2m}\right) = \rho(S) - \sum_{u \in S, v \in S} \frac{d_u d_v}{2m} \tag{3}$$

### 5.4.6 Cohesion

Cohesion [13] is a recently proposed metric for community detection problems. It generalizes the notion of cuts. Cuts are edges that cross the boundary of a vertex set. Cohesion measures the triangles that cross the boundary of a vertex set. We first define the set of triangles. Given three vertex sets $S_1$, $S_2$ and $S_3$, the triangles spanned by the three sets are:

$$\Delta(S_1, S_2, S_3) = \{(u,v,w) | u \in S_1, v \in S_2, w \in S_3, (u,v) \in E, (v,w) \in E, (w,u) \in E\}$$

Note that we use $(u,v,w)$ to denote an unordered triplet. The inner triangles of a set of vertices $S$ are $\Delta_i(S) = \Delta(S,S,S)$. The boundary (outer) triangles are $\Delta_o(S) = \Delta(S,S,V-S)$. The original cohesion is defined as:

$$\mathrm{Cohesion}(S) = \frac{|\Delta_i(S)|}{\binom{|S|}{3}} \times \frac{|\Delta_i(S)|}{|\Delta_i(S)| + |\Delta_o(S)|}$$

---

[3]. This is an after thought when I finish most of the paper. By adding the $D$ in quadratic form section, all VIES are encoded by quadratic forms. So quadratic form turns out to be the most general in this series. This is the same observation made by prof. Lau.

The first term considers the density of the group of vertices in terms of triangles. The second term measures how isolated the group of vertices are. Fig. 1 in [13] is a good illustration of why cut fails but cohesion succeeds in capturing certain community structures.

During the sampling procedure, the number of triangles are deemed to decrease and we can quantify this. Suppose the edge sampling probability is $p$. Using the linearity of expectations

$$\begin{aligned} \text{E}[\# \text{ Triangles}] &= \sum_{(u,v,w) \in \Delta(S_1, S_2, S_3)} \Pr\{(u,v)\} \times \Pr\{(v,w)\} \times \Pr\{(w,u)\} \\ &= |\Delta(S_1, S_2, S_3)| \, p^3 \end{aligned}$$

The triangle density will also decrease. Maybe we can fix this by up-weighting edges. Leave it to future discussions. The second term is homogeneous. That is, after sampling, the value stays the same in terms of expectation. Note that besides $\Delta_i$ and $\Delta_o$ defined above, there is another type: $\Delta(S, V-S, V-S)$. Let $\Delta(S)$ be all the triangles related with $S$:

$$\Delta(S) = \Delta_i(S) + \Delta_o(S) + \Delta(S, V-S, V-S)$$

We can define, Triangle Cut (TCut), a similar quantity to NCut

$$\text{TCut}(S) = \frac{|\Delta_o(S)|}{|\Delta(S)|}$$

and Triangle Association (TAssoc)

$$\text{TAssoc} = \frac{|\Delta_i(S)|}{|\Delta(S)|}$$

### 5.4.7 Properties with Different Vertex Set

The above mentioned properties are all defined with full vertex set, i.e. after sampling, $V_s = V$. If the vertex sets are different, it's very hard to say, for example, what means "a cut is preserved". Other properties like quadratic forms also suffer from this problem. Towards this end, one will find those properties discussed in this section are always studied along with edge sampling.

We want to extend the definitions of those properties so that they are meaningful after vertex sampling. [44] defines such an objective. Suppose we have a set of terminal vertices $K$. After graph transformation, the resulting graph preserves all min-cuts between $U$ and $K - U$, where $U \subseteq K$.

How to come up with meaningful counterparts of the properties discussed in this section is still a problem.

## 5.5 Distance Metrics for Properties

Note that there are basically two types of properties mentioned above. One type is vector or distribution in nature, e.g. Degree Distribution, Radius. The other type is a digest of the distribution property by taking max, min, or average, e.g. Average Degree Distribution, Diameter (max of Radius).

Given the original graph $G$ and sampled graph $G_s$, we want to know how far is $f(G_s)$ from $f(G)$. The second type is just a scalar and the common measure of the quality of estimation is by Normalized Root Mean Square Error: (e.g. [29][63][51])

$$\text{NRMSE}(\theta, \hat{\theta}) = \frac{\sqrt{\text{E}[(\hat{\theta} - \theta)^2]}}{\theta}$$

For the first type, some distance metrics are used. Let $p$ be the distribution derived from original graph and $q$ be the distribution derived from sampled graph. Suppose they are defined on $\Omega$, one can measure the distance in the following ways:

- Total variation distance (e.g. [51]) measures all the difference between two distributions:

$$\text{TV}(p, q) = \max_{A \subseteq \Omega} |p(A) - q(A)| = \frac{1}{2} \sum_{x \in \Omega} |p(x) - q(x)|$$

- Kullback-Leibler (KL) divergence captures the difference of the two distributions accounting for the bulk of the distributions:

$$\mathrm{KL}(p, q) = \sum_{x \in \Omega} p(x) \ln \frac{p(x)}{q(x)}$$

- Kolmogorov-Smirnov (KS) statistic (e.g [2]) captures the maximum vertical distance of the c.d.f of the two distributions. When $\Omega = \mathbb{R}$, it can be defined as:

$$\mathrm{KS}(p, q) = \max_{x} \left| \sum_{\xi \leqslant x} p(\xi) - \sum_{\xi \leqslant x} q(\xi) \right|$$

# 6 Property Preservation/ Estimation Results

In this section we discuss some ad hoc results of property preservation or property estimation. Although the two objectives are initially different, their results can transform to each other as discussed before. This section is still under development. We have markers for sections coming in future versions of this survey. Interested readers can come back to `arxiv` to retrieve them.

## 6.1 Network Size Estimation

Network size estimation may be the most direct objective of graph sampling. However, we did not find many works on this direction. There were previous works for population estimation (Section 6.1.3), but they are not tuned for graphs or not designed to utilize graph structures.

### 6.1.1 Graph Identities on NS, GD and AD

We use $n = |V|$ and $m = |E|$ to describe the size of a network. They are related with graph density $\rho$[4] and average degree $\langle d \rangle$. The following graph identities are useful

$$\rho = \frac{2m}{n(n-1)} \tag{4}$$

$$\langle d \rangle = \frac{2m}{n} \tag{5}$$

$$n = \frac{\langle d \rangle}{\rho} + 1 \tag{6}$$

One may estimate some of those quantities and use the above identities to estimate rest ones. For example, the estimation of average degree is well understood (Section 6.1.2). There are also many classical methods for population estimation, leading to many (may not be so accurate) estimators for $n$ (Section 6.1.3). Combining the two quantities, we can estimate $m$ and $\rho$ using those identities. [28] takes a different approach: estimate $\rho$ and $\langle d \rangle$; use graph identities to get $n$ and $m$.

### 6.1.2 Average Degree

One can apply the analysis for degree distribution to see that average degree is not preserved by VS and ES. However, by VSN (practical for OSN crawling), one can obtain the neighbourhood. Averaging over directly crawled vertices yields an unbiased estimator for average degree of $G$:

$$\widehat{\langle d \rangle} = \frac{1}{|\tilde{V}_s|} \sum_{v \in \tilde{V}_s} d(v) \tag{7}$$

If we use RW, it's known that the samples are biased towards high degree vertices. We can use RDS to correct it. Simply do RW on the graph and substitute $g(v) = d(v)$ in Eq. 2:

$$\widehat{\langle d \rangle} = \frac{|V_s|}{\sum_{v \in V_s} \frac{1}{d(v)}} \tag{8}$$

---

4. $\rho$ is overloaded in the discussion of association.

This is to say, we use the harmonic mean of the degrees of our sampled vertices $V_s$ as the estimator for the mean degree of the graph.

### 6.1.3 Population Estimation

We digest classical population estimation methods from the pointers in [28]. Interested readers can see reference therein. The population estimation problem assumes the existence of vertex samplers (repetitions are allowed). Both uniform and non-uniform sample are workable, as long as we know the distribution. We discuss some classical methods using uniform samples and the estimators on non-uniform sample can be designed similarly: (using Hansen-Hurwitz estimator [19])

- Capture-recapture. Suppose we have $S_1$ and $S_2$ being two independent samples without replacement (only retaining unique elements). The expected number of intersection is

$$\mathrm{E}[|S_1 \cap S_2|] = \sum_{v \in S_1} \Pr\{v \in S_2\} = \frac{|S_1||S_2|}{|V|}$$

  So one estimator is

$$\hat{n} = \frac{|S_1||S_2|}{|S_1 \cap S_2|}$$

- Unique element counting. Let $S$ be the sample with replacement (with repetitions) and $S^{\mathrm{uniq}}$ be the unique elements from $S$. Counting the unique elements are just a balls-and-bins problem: throw $|S|$ balls into $n$ bins; how many bins are occupied?

$$\mathrm{E}[|S^{\mathrm{uniq}}|] = \sum_{i=1}^{n} \Pr\{v_i \in S\} = \sum_{i=1}^{n} \left(1 - \left(1 - \frac{1}{n}\right)^{|S|}\right) \approx n\left(1 - e^{-\frac{|S|}{n}}\right) \quad (9)$$

  One can solve $n$ using observed quantities as an estimator.

- Collision counting. Pick two vertices from a set of sample $S = \{s_1, s_2..., s_{|S|}\}$. If they are the same vertex, we call it a collision. The expected number of collisions are:

$$\mathrm{E}[N] = \sum_{i<j} \Pr\{s_i = s_j\} = \binom{|S|}{2}\frac{1}{n}$$

  One can obtain the collision counting estimator:

$$\hat{n} = \binom{|S|}{2}/N$$

How to quantify the accuracy of those estimators is of its own research interest in population estimation.

### 6.1.4 Density Estimation

The method [28] uses to estimate density is very similar to collision counting estimation for population size. The graph density can be interpreted as the probability that a randomly chosen pair of vertices are adjacent. Denote the sampled vertices (with replacement) by $S$ and number of adjacent pairs in $S$ by $N$, namely $N = \sum_{i<j} \mathrm{I}[(s_i, s_j) \in E]$. We have the following relation:

$$\mathrm{E}[N] = \sum_{i<j} \Pr\{(s_i, s_j) \in E\} = \binom{|S|}{2}\rho$$

Similar estimator can be designed if vertex sampling is non-uniform. See [28] for more details.

## 6.2 Full Graph Observation

In graph crawling, the first question to ask is: How many vertices/ edges do we need to sample in order to observe a considerable portion (e.g. $1 - \varepsilon$ fraction) of the graph?

### 6.2.1 Vertex Sampling

There are two scenarios:

- If we can only do VS with replacements. This is a coupon collector problem and we know the number of samples we need in order to observe the whole graph is $\Theta(n \ln n + c\, n)$. It has a sharp threshold behaviour at this value. In order to observe $1-\varepsilon$ fraction of the vertices, the number of samples $s$ must satisfy: (using Eq. 9)

$$n\left(1 - e^{-\frac{s}{n}}\right) \geqslant (1-\varepsilon)n$$

which gives:

$$s \geqslant n \ln \frac{1}{\varepsilon}$$

- If we can do VS without replacements, it becomes trivial: $s \geqslant (1-\varepsilon)n$.

### 6.2.2 Uniform Vertex Sampling with Neighbourhood

VSN makes the problem more interesting and is the usual case of OSN crawling. We only consider the scenario without replacements, because one can easily remove duplicates during the crawling.

If the min degree of the graph is $d_{\min}$, we expect VSN to act $d_{\min}$ times more efficient than VS. However, the neighbourhood of vertices sampled later may have larger probability to have already been observed in previous crawling. Denote the sampled set by $S$ (with repetition). The number of new vertices introduced by one crawl when there are already $N$ unique vertices (including the neighbouring vertices of $S$) is:

$$\begin{aligned}
\mathrm{E}[\#\,\mathrm{new}|N] &= \sum_{v \in V-S} \Pr\{\text{select } v\} \sum_{u \in N(v)} \Pr\{u \notin S\} \\
&= \sum_{v \in V-S} \Pr\{\text{select } v\} d(v)\left(1 - \frac{N}{n}\right) \\
&\approx \sum_{v \in V} \Pr\{\text{select } v\} d(v)\left(1 - \frac{N}{n}\right) \\
&= \langle d \rangle \left(1 - \frac{N}{n}\right)
\end{aligned}$$

The approximation holds because by uniform VS, $S$, $V-S$ and $V$ should have approximately same degree distribution (Section 6.3.3). Denote the number of unique vertices at step $i$ by $N_i$. The expected number of unique vertices after $t$ steps can be calculated as:

$$\begin{aligned}
\mathrm{E}[N_t] &= \sum_x (\mathrm{E}[\#\,\mathrm{new}|x] + x)\Pr\{N_{t-1} = x\} \\
&= \sum_x \left(\langle d \rangle + \left(1 - \frac{\langle d \rangle}{n}\right)x\right) \Pr\{N_{t-1} = x\} \\
&= \langle d \rangle + \left(1 - \frac{\langle d \rangle}{n}\right) \mathrm{E}[N_{t-1}]
\end{aligned}$$

The boundary condition is $\mathrm{E}[N_1] = \langle d \rangle$. So

$$\begin{aligned}
\mathrm{E}[N_t] &= \langle d \rangle + \left(1 - \frac{\langle d \rangle}{n}\right)\langle d \rangle + \ldots \left(1 - \frac{\langle d \rangle}{n}\right)^{t-1} \mathrm{E}[N_1] \\
&= \langle d \rangle \frac{1 - \left(1 - \frac{\langle d \rangle}{n}\right)^t}{1 - \left(1 - \frac{\langle d \rangle}{n}\right)} \\
&= n\left(1 - \left(1 - \frac{\langle d \rangle}{n}\right)^t\right)
\end{aligned}$$

In order to observe $1-\varepsilon$ fraction of the vertices, we need to have

$$t \geqslant \frac{n}{\langle d \rangle} \ln \frac{1}{\varepsilon}$$

which is $\langle d \rangle$ times more efficient than VS.

### 6.2.3 Non-uniform Vertex Sampling with Neighbourhood

The uniform VSN is $\langle d \rangle$ times more efficient than VS, because sampling a vertex can give us some neighbours. However, this approach is not optimal. We can try to sample higher degree vertices first which potentially bring us more new vertices in one crawl. The number of new vertices depends largely on what we have crawled and the argument is not easy. We can only provide a loose lower bound:

$$t \geqslant \frac{n}{d_{\max}} \ln \frac{1}{\varepsilon}$$

### 6.2.4 Edge Sampling

In the usual sense, ES only selects an edge and its two endpoints. If we only look at one vertex, it degrades to VS with `degree proportional distribution`. The efficiency will not exceed ordinary uniform VS. In order to give a loose lower bound, we can treat the two endpoints as independently picked from the graph. That is:

$$t \geqslant \frac{n}{2} \ln \frac{1}{\varepsilon}$$

To give a tighter bound, we can consider the smallest degree vertices. It will have at least $\frac{1}{2m}$ to be picked at each round. So we have:

$$t \geqslant m \ln \frac{1}{\varepsilon}$$

### 6.2.5 Traversal Based Sampling

Most TBS are variants of RW. So this becomes the question of finding the cover time of a Markov chain. Before knowing detailed information about the graph, one can apply a simple bound, $O(n\,m)$ [33, L6] if it is connected. A tighter version for general graph is [33, L6]:

$$m\,R(G) \leqslant C(G) \leqslant 2e^3\,m\,R(G) \ln n + n$$

where $R(G) = \max R_{u,v}$ and $R_{u,v}$ is the effective resistance of the graph. Many social graphs turn out to be small world graphs, that is, the diameter is $O(\ln n)$. Find such a pair of vertices $u, v$ who has this maximum shortest path length. We apply Rayleigh's monotonicity principle of resistance: After removing all edges not on the shortest path, the resistance can only increase. So we know $R_{u,v} = O(\ln n)$. We can bound the cover time:

$$C(G) = O(2e^3 m \ln^2 n + n)$$

This is not too much worse than VS with replacement. When one can not enumerate ID space of an OSN, RW may also be effective.

Note that cover time guarantees all vertices are visited. If we only asks $1 - \varepsilon$ fraction, similar analysis can be adapted. First, the maximum hitting time is $H = 2m \ln n$ for our small-world graph ($2m\,R(G)$, [33, L6]), regardless of the starting vertex. After $2H$ steps, the probability that one vertex is not visited is $1/2$ by Markov's inequality. After $t \times 2H$ steps, the probability that one vertex is not visited is $(1/2)^t$. We use the linearity of expectation to compute the number of visited vertices:

$$\mathrm{E}[N] = \sum_{v \in V} \Pr\left\{v \text{ visited}\right\} = n\left(1 - \left(\frac{1}{2}\right)^t\right)$$

To make it larger than $1 - \varepsilon$, we have $t \geqslant \log_2 (1/\varepsilon)$. So it takes $2m \ln n \log_2 (1/\varepsilon)$ time to observe $1 - \varepsilon$ fraction of all the vertices.

## 6.3 Degree Distribution

Denote the probability density function of degree distribution for the original graph as $p(k)$. To obtain the degree distribution of the sampled graph, we simply calculate every point of $p_s(k)$, i.e. $p_s(k) = \Pr\{d_{G_s}(X) = k\}$, where $X \in V_s$ is a random vertex picked from the sampled graph.

### 6.3.1 Vertex Sampling

To obtain a degree $k$ vertex in $G_s$, we basically need two phases: 1) one $v \in V$ with $\Pr\{d_G(v) \geqslant k\}$ is picked; 2) $v$ and $k-1$ of its neighbours are preserved in the sampling. That is: ([35])

$$\begin{aligned} p_s(k) &= \Pr\{d_{G_s}(X) = k\} \\ &= \sum_{j \geqslant k} \Pr\{d_{G_s}(X) = k | d_G(X) = j\} \Pr\{d_G(X) = j\} \\ &= \sum_{j \geqslant k} \frac{\binom{j}{k}\binom{n-1-j}{n_s-1-k}}{\binom{n-1}{n_s-1}} p(j) \end{aligned} \qquad (10)$$

If we assume $n_s$ is not fixed but we only sample every vertex independently with probability $\alpha = \tilde{n}_s/n$, where $\tilde{n}_s$ is the intended number of vertices in the sampled graph. Then the degree distribution becomes: ([59], [35]*)

$$\begin{aligned} p_s(k) &= \sum_{j \geqslant k} \Pr\{d_{G_s}(X) = k | d_G(X) = j\} \Pr\{d_G(X) = j\} \\ &= \sum_{j \geqslant k} \binom{j}{k} \alpha^k (1-\alpha)^{j-k} p(j) \end{aligned} \qquad (11)$$

When $n \to \infty$ but the ratio $\alpha = n_s/n$ is fixed, Eq. 10 is approximately equivalent to Eq. 11:

$$\begin{aligned} \frac{\binom{n-1-j}{n_s-1-k}}{\binom{n-1}{n_s-1}} &= \frac{\frac{(n-1-j)!}{(n-j-n_s+k)!(n_s-1-k)!}}{\frac{(n-1)!}{(n-n_s)!(n_s-1)!}} \\ &= \frac{(n-n_s)!}{(n-j-n_s+k)!} \frac{(n-1-j)!}{(n-1)!} \frac{(n_s-1)!}{(n_s-1-k)!} \\ &\approx (n-n_s)^{j-k}(n-1)^{-j}(n_s-1)^k \\ &\approx (1-\alpha)^{j-k}\alpha^k \end{aligned}$$

### 6.3.2 Edge Sampling

ES turns out to give the same result as Eq. 11 ([35]). The first equality holds regardless of the sampling technique that is used. To argue the conditional probability in the summation, note that "$k$ out of $j$ neighbours are sampled" is equivalent to "$k$ out of $j$ incident edges are sampled", thus giving the same result.

Eq. 11 only confirms one intuitive conclusion: vertices on $G_s$ in general have lower degree than vertices in $G$. Except for this, it does not lead to useful application. As noted in Section 5.4.1, cut subsumes neighbourhood. Using this fact, we can apply the ES for cut to find an estimator of degree. More concretely, if the sample probability of edges is $p$, we set the weight of edges in $E_s$ to be $1/p$. We calculate the (weighted) degree distribution of $G_s$, which is an estimator for the degree distribution of $G$, with bounded error.

### 6.3.3 Vertex Sampling with Neighbourhood

VSN will give precise degree for $v \in \tilde{V}_s$ (not $V_s$). So a straightforward estimator for degree distribution is

$$p_s(k) = \frac{1}{|\tilde{V}_s|} \sum_{v \in \tilde{V}_s} \mathrm{I}[d(v) = k]$$

If we treat $\tilde{V}_s$ as a r.v., it can be shown that

$$\begin{aligned} p_s(k) &= \Pr\{d_{G_s}(X) = k | X \in \tilde{V}_s\} \\ &= \Pr\{d_{G_s}(X) = k | X \in \tilde{V}_s, d_G(X) = k\} \Pr\{d_G(X) = k | X \in \tilde{V}_s\} \\ &= 1 \times \Pr\{d_G(X) = k\} \\ &= p(k) \end{aligned}$$

So the two distributions are asymptotically the same. If degree is obtained from a vertex label instead of topology, the same result holds.

### 6.3.4 Edge Sampling with Vertex Label

In the normal ES, we only observe a part of the graph and degrees should be underestimated. If we further assume the availability of vertex label for all vertices in $V_s$ and degree is presented as one vertex label, we can get similar result like VSN.

If we treat $\tilde{V}_s$ as a r.v., it can be shown that: ([41]* [5])

$$\begin{aligned}
\Pr\{d_{G_s}(X) = k | X \in \tilde{V}_s\} &= \Pr\{d_{G_s}(X) = k | X \in \tilde{V}_s, d_G(X) = k\} \Pr\{d_G(X) = k | X \in \tilde{V}_s\} \\
&= 1 \times \frac{\Pr\{X \in \tilde{V}_s | d_G(X) = k\} \Pr\{d_G(X) = k\}}{\Pr\{X \in \tilde{V}_s\}} \\
&= \frac{\frac{k}{2m} \times p(k)}{\Pr\{X \in \tilde{V}_s\}} \\
&\propto k\, p(k)
\end{aligned}$$

That is to say, the degree distribution on $V_s$ is biased towards high degree edges if we use ES. We can correct the bias by the following estimator:

$$p_s(k) = \frac{1}{Zk} \sum_{v \in \tilde{V}_s} \mathrm{I}[d(v) = k]$$

where $Z$ is a normalization constant. This $p_s(k)$ will be asymptotically the same as $p(k)$.

### 6.3.5 BFS

[30] shows that BFS has a bias towards high degree vertices.

**This section will be provided in a future version. To checkout our current progress and previews, please contact the authors.**

### 6.3.6 Results for Certain Type of Graphs

Note that the above discussion does not assume a certain type of graph. In the survey, we found the results for certain type of graphs:

- [60] provides a study of scale-free (power-law) graphs, where $p(k) \propto k^{-\gamma}$. There is an exact analytic expression for $\gamma = 2$.
- On Erdos-Renyi random graphs, one can see chapter 2 of [10].

## 6.4 Minimum Cut

If the min-cut of a graph is preserved on $G_s$, a min-cut algorithm (e.g. runs in $O(n_s^3)$) will of course output it. In this section, we study the sample and contract approach. [43] presents an extreme of such sampling approach, i.e. sample and contract until $V_s = 2$. Now we want to investigate how likely an early stopping will preserve this property, i.e. min-cut. Stopping earlier and running min-cut for $O(n_s^3)$ turns out to be an improvement for both algorithm, i.e. $O(n^4)$ random contraction algorithm and $O(n^3)$ best deterministic min-cut algorithm.

### 6.4.1 Edge Sampling with Contraction

Although we do not give the best randomized algorithm in this section (see L1 of [33] for some pointers), it suffices to show the flavour of our core objective: Use simple algorithms (e.g. ES, VS, ESC, VSC) to reduce the graph with the hope of preserving certain property of the graph; Then we run algorithms depending on this property to obtain (approximately) the same result.

**Framework**

Our framework is simple:

1. Pick an arbitrary edge and contract it. Stop until there are only $r$ vertices. Get $G_s$.

---

5. "edge sampling" mentioned in some literatures are actually "edge sampling with vertex label". This is the common confusion part because you sometimes see degree is preserved (or can be estimated) but sometimes not. The assumption of vertex label availability after ES is valid in some scenarios. For example, activity (e.g. retweet) edge on OSN often contains such information (available via API): tweet content, tweet-er info, retweet-er info. One can get some vertex label from tweet-er or retweet-er info, like follower count, followee count, etc. (holds for SinaWeibo)

2. Run any deterministic min-cut algorithm on $G_s$ and record the cut.

3. Repeat the above experiments $\Theta(n^2/r^2)$ (calculated later) times. Output the minimum one of all those experiments.

It can be shown that the min-cut is output w.h.p..

**Previous Results**

The best deterministic algorithm for min-cut is know to be $O(n^3)$. [43] (Ch. 1) presented a simple randomized algorithm, which runs in $O(n^4)$ time. That algorithm is a special case of our framework by substituting $r=2$. Since only two vertices are left in $G_s$, the deterministic min-cut algorithm is trivial (there is only one cut). However, this approach has very high failure probability, so we have to repeat the experiment for $\Theta(n^2)$ times in order to get the real min-cut w.h.p.. Combining with the contraction cost, it gives an $O(n^4)$ algorithm.

**Our results**

The analysis is very similar to that of [43] (Ch. 1), except that we stop early and defer the determination of best $r$.

Denote the event that the edge contracted at $i$-th iteration is not in $C$ (the min-cut) by $E_i$. Let $F_i = \cap_{j=1}^i E_j$. It can be shown that $P(E_1 = F_1) \geqslant 1 - \frac{2}{n}$ and $P(E_i|F_{i-1}) \geqslant 1 - \frac{k}{k(n-i+1)/2} = 1 - \frac{2}{n-i+1}$ ($k = |C|$). The probability that the min-cut property is preserved after $n-r$ round is:

$$\begin{aligned} P(F_{n-r}) &= P(E_{n-r}|F_{n-r-1})...P(E_2|F_1)P(F_1) \\ &\geqslant \prod_{i=1}^{n-r}\left(\frac{n-i-1}{n-i+1}\right) \\ &= \frac{(n-1-(n-r))(n-1-(n-r-1))}{n(n-1)} \\ &= \frac{(r-1)r}{n(n-1)} \end{aligned}$$

Consider $r = r(n)$ as a function of $n$, this probability is $\Theta\left(\left(\frac{r(n)}{n}\right)^2\right)$ and by repeating the experiment for $\tilde{\Theta}\left(\left(\frac{n}{r(n)}\right)^2\right)$ times, we are certainly to preserve the min-cut in at one experiment (fail probability smaller than $O\left(\frac{1}{\text{poly}(n)}\right)$).

Running min-cut algorithm on the sampled graph takes $O(r^3)$. The contraction procedure can be implemented in $O(n^2)$ time. So the total complexity for this algorithm is:

$$O\left(\frac{n^2}{r^2}(n^2+r^3)\right)$$

Let $r(n) = n^c$, this becomes

$$n^{2-2c}(n^2+n^{3c}) = n^{4-2c} + n^{2+c} \geqslant 2\sqrt{n^{4-2c}n^{2+c}}.$$

The "=" can be achieved with $c = \frac{2}{3}$. Plugging it back, we get a min-cut algorithm of complexity $O(n^{8/3}) < O(n^3)$. This is better than running the min-cut on original graph.

There are some future extensions to make the argument in this section tighter:

- Note that the deterministic min-cut can achieve $O(n\,m)$ which is less than $O(n^3)$ on sparse graphs. One can extend the argument in this section by taking the relationship of $m$ and $n$ into consideration, i.e. $m = n^b$ for a known $b$.
- One contraction down to $r$ vertices only takes $O(n(n-r))$ time, which is slightly smaller in implementation. Although it does not change the order in theoretical analysis, it may benefit implementations for not so large $n$ and $m$.

### 6.4.2 Vertex Sampling with Contraction

Similar idea can be applied to vertex sampling with contraction. Denote the set of vertices incident to $C$ by $V_C$. We know that $|V_C| \leqslant 2|C| = 2k$. The goal is to calculate the probability that such $2k$ vertices survive the VSC process. Through similar derivation, we can get the lower bound of this probability: (let $r(n) = n^c$)

$$\frac{(r-2k)^{2k}}{n^{2k}} \approx n^{(c-1)2k}$$

The optimal choice for $c$ also comes at $c = \frac{8}{3}$. However, the complexity is $O(n^{8k/3})$, much larger than that of the ESC.

### 6.4.3 ESC Approximation

The above ESC algorithm aims at exact calculation of min-cut so repeated experiments are required to increase the success probability. If we only want to approximate min-cut, maybe we can get more efficient algorithms. That is, after running ESC, with probability $1 - \varepsilon$, at least one cut with size smaller than $(1+\delta)|C|$ is preserved.

**This section will be provided in a future version. To checkout our current progress and previews, please contact the authors.**

## 6.5 Cut

### 6.5.1 Uniform Edge Sampling

There is an edge sampling algorithm to preserve cuts of all subsets of vertices. We present a sketch here. Interested readers can see [33, L3] for more details. The algorithm is simple: perform ES with probability $p$; $\forall e \in E_s$, set weight $w(e) = \frac{1}{p}$. It can be shown that with some subtle conditions, $G_s$ preserves all the cuts.

Denote the weight of a cut by $w_G(\delta_G(S)) = \sum_{e \in \delta_G(S)} w_G(e)$, where $w_G(e)$ is the weight of edge $e$ in graph $G$. In an unweighted graph, $w(e) = 1$. We denote $H = G_s$ to make it uncluttered. The technical expression for "preserving all cuts" is:

$$(1 - \varepsilon) w_G(\delta_G(S)) \leqslant w_H(\delta_H(S)) \leqslant (1 + \varepsilon) w_G(\delta_G(S)) \; \forall S \in V$$

In the discussion, we also call $\delta(S)$ as "cut $S$". Since the number of edges of cut $S$ obeys binomial distribution Binomial$(|S|, p)$ and the weight of remaining edges is set to $1/p$, the expected size of $\delta(S)$ and weight of $\delta(S)$ are:

$$\begin{aligned} \mathrm{E}[|\delta_H(S)|] &= |\delta_G(S)| \times p = p|\delta_G(S)| \\ \mathrm{E}[w_H(\delta_H(S))] &= |\delta_G(S)| \times p \times \frac{1}{p} = |\delta_G(S)| \end{aligned}$$

which is the same as the weight in original graph. We use chernoff bound (Appendix A) to calculate the deviation probability of the size of $\delta(S)$:

$$\Pr\left\{||\delta_H(S)| - p|\delta_G(S)|| \geqslant \varepsilon p|\delta_G(S)|\right\} \leqslant 2 e^{-p|\delta_G(S)|\varepsilon^2/3}$$

Denote the event that cut $S$ is preserved by $E_S$ and the fail probability is $\Pr\{\bar{E}_S\} \leqslant 2 e^{-p|\delta_G(S)|\varepsilon^2/3}$. We want to apply union bound on all $S \in V$. It can be shown (Appendix A) that there are no more than $n^{2\alpha}$ cuts of size larger than $\alpha c$, where $c = \Omega(\ln n)$ is the minimum cut size. Then we get:

$$\begin{aligned} \Pr\left\{\bigcup_{S \in V} \bar{E}_S\right\} &\leqslant \sum_{S \in V} \Pr\{\bar{E}_S\} \\ &\leqslant \sum_{\alpha \geqslant 1} \sum_{S \in V, |\delta_G(S)| = \alpha c} \Pr\{\bar{E}_S\} \\ &\leqslant \sum_{\alpha \geqslant 1} \sum_{S \in V, |\delta_G(S)| = \alpha c} 2 e^{-p\alpha c \varepsilon^2/3} \\ &\leqslant \sum_{\alpha \geqslant 1} \sum_{S \in V, |\delta_G(S)| \leqslant \alpha c} 2 e^{-p\alpha c \varepsilon^2/3} \\ &\leqslant \sum_{\alpha \geqslant 1} n^{2\alpha} \times 2 e^{-p\alpha c \varepsilon^2/3} \end{aligned}$$

If we let $p = \frac{3(d+2)\ln n}{\varepsilon^2 c}$,

$$\Pr\left\{\bigcup_{S \in V} \bar{E}_S\right\} \leqslant \sum_{\alpha \geqslant 1} n^{2\alpha} \times 2 \times n^{-\alpha(d+2)} = 2 \times \sum_{\alpha \geqslant 1} n^{-\alpha d} = O(n^{-d})$$

So all the cuts are preserved w.h.p..

### 6.5.2 Non-uniform Edge Sampling Using Edge Strong Connectivity

[8] proposes non-uniform edge sampling technique in order to relax the constraint that $c = \Omega(\ln n)$. The idea is to sample edges in sparse cut with higher probability and lower weight, and sample edges in dense cut with lower probability and higher weight. It can effectively reduce the number of edges to $O(n \ln n)$. See also [33, L11 of 2011] for a digest.

## 6.6 RCut, NCut, Assoc, RAssoc, NAssoc, Volume

This section will be provided in a future version. To checkout our current progress and previews, please contact the authors.

## 6.7 Modularity

This section will be provided in a future version. To checkout our current progress and previews, please contact the authors.

## 6.8 Cohesion

This section will be provided in a future version. To checkout our current progress and previews, please contact the authors.

## 6.9 Quadratic Forms

### 6.9.1 Non-uniform Edge Sampling Using Effective Resistance

[32, L12] presents one non-uniform edges sampling to preserve quadratic forms of Laplacian. The algorithm is:

- Initialize $H$ as an empty graph.
- For $1 \leq i \leq q$
    - Pick random edge $e$ with probability $p_e$.
    - Increase the weight of $e$ in $H$ by $\frac{w_e}{q p_e}$.

It can be shown that by setting

$$p_e = \frac{w_e R_e}{\sum_e w_e R_e}, q = O(n \log n / \varepsilon^2)$$

where $R_e$ denotes the effective resistance of $e$, we can get an $\varepsilon$-approximation, namely:

$$(1-\epsilon) x^\mathrm{T} L_G x \leq x^\mathrm{T} L_H x \leq (1+\epsilon) x^\mathrm{T} L_G x, \forall x \in \mathbb{R}^n$$

We'll omit the detailed discussion but present some notes here for operational convenience

- The effective resistance between vertex $a$ and $b$ is defined as the $v(a) - v(b)$ when one Ampere current is injected into $a$ and removed from $b$, where $v(a)$ denotes the voltage of $a$. Let $w(e)$ be the (current) conductance (inverse resistance) of edge $e$, i.e. $A_{u,v} = w(e)$ for $e = (u, v)$. In this way, effective resistance can be expressed as

$$(\chi_a - \chi_b)^\mathrm{T} L_G^{-1} (\chi_a - \chi_b) \qquad (12)$$

  where $L_G^{-1}$ denotes the pseudo inverse and $\chi_a$ is the characteristic vector for a single vertex.

- In order to run this algorithm, one need to evaluate Eq. 12. This is to either solve the pseudo inverse or to solve a Laplacian system, i.e. $L x = b$. The state-of-the-art linear solver for Laplacian system involves this spectral sparsification process. The recursive construction of linear time Laplacian solver and spectral graph sparsifier is too complex to present here. See also [32, L11-L13] for details.

## 6.10 Shortest Path Length

This section will be provided in a future version. To checkout our current progress and previews, please contact the authors.

# 7 Conclusion and Future Works

Graph sampling is a well motivated topic and has important applications in all kinds of research fields. There are not so many theoretical analysis on property preservation or estimation. For numerical evaluation, people used different criteria, algorithms, graphs and properties. There are many ad hoc results of property preservation. Some future works are:

- A comprehensive numerical evaluation of the property preservation result of different sampling algorithms. One can see that there is a large vacancy in the combination, i.e. properties (Section 5.2) × algorithms (Section 3.3) × criteria (Section 5.5) × graph models (Section 3.2). All previous works did part of them. A comprehensive and neutral numerical evaluation can hint the possible directions of future theoretical studies.

- There is a lack of evaluation on synthesized data set. While evaluation on real networks give convincing results of whether an algorithm is applicable or not, the evaluation on synthesized data set (using certain graph generation model) may give people some insights between algorithms and properties.

- There is only a few theoretical studies. One can see that the most well understood property is degree distribution and its derivatives. It's non-trivial to analytically quantify some of the properties. However, some easier ones are tractable.

- TBS techniques are quite popular in recent literatures. They are also promising practical techniques for large graph crawling. However, most works do not provide theoretical analysis of the variance and efficiency of those estimators. The theoretical analysis like [45][34] can lay a more solid foundation for future application.

- **Problem oriented property preservation (POPP)** is at its infancy. As we mentioned before, in many research fields, a family of algorithms may aim at optimizing a certain property defined on a graph. If $G_s$ exhibits similar properties as $G$, we may expect those algorithms also work on $G_s$. This gives a way to probabilistically accelerate a class of algorithms (maybe with bounded loss of optimality). The simple min-cut improvement made in this survey (Section 6.4) shows a flavour of POPP. Since POPP is tightly binded with the graph algorithms, more further investigations can be done.

# Appendix A  Useful Probability Results

**Theorem 1.** *(Chernoff Bounds) Consider a heterogeneous coin flipping setting: $X_1, X_2, ... X_n$ with head probability $p_1, p_2, ... p_n$. Let $X = \sum_{i=1}^{n} X_i$ and $\mu = \mathrm{E}[X] = \sum_{i=1}^{n} p_i$. We have the following conclusions: ([33] L3)*

1. *$\mathrm{E}[e^{tX}] \leqslant e^{\mu(e^t - 1)}$ is the general result.*
2. *for $\delta > 0$, $\Pr\{X \geqslant (1+\delta)\mu\} < \left(\frac{e^\delta}{(1+\delta)^{1+\delta}}\right)^\mu$. This is the strongest among the following specific results.*
3. *for $0 < \delta < 1$, $\Pr\{X \geqslant (1+\delta)\mu\} < e^{-\delta^2\mu/3}$. This is more practical*
4. *for $R \geqslant 6\mu$, $\Pr\{X \geqslant R\} \leqslant 2^{-R}$.*
5. *for $0 < \delta < 1$, $\Pr\{X \leqslant (1-\delta)\mu\} \leqslant \left(\frac{e^{-\delta}}{(1-\delta)^{1-\delta}}\right)^\mu$*
6. *for $0 < \delta < 1$, $\Pr\{X \leqslant (1-\delta)\mu\} \leqslant e^{-\mu\delta^2/2}$.*
7. *for $0 < \delta < 1$, $\Pr\{|X - \mu| \geqslant \delta\mu\} \leqslant e^{-\mu\delta^2/3}$. This is a frequently used form. It's the double sided version derived from (3) and (6).*
8. *Same results hold if $X_i \sim U[0,1]$ with $\mathrm{E}[X_i] = p_i$. (Hoeffding's extension).*

**Lemma 2.** *There are at most $n^{2\alpha}$ cuts with $\alpha c$ edges, where c is the min-cut value. ([33] HW1)*

# Appendix B  Long Derivations

# Appendix C  Other Works

There are many works related to graph sampling but not systematically discussed in this paper. In this section, we briefly mention some of those works found during this survey. After proper organization, they will be put in the future version of this survey.

[53] studies the sampling on directed graphs. [5] addressed the ability of streaming sampling of a large number of edges from social activity graphs. [3] proposes a graph induction step after ES, i.e. after obtaining $V_s$ and $E_s$ in our terminology, edges between $V_s$ are also added. The use case is when (activity) edges come in order of time and we want to sample the topology in one pass o the data. [2] is one Problem Oriented Property Preservation (PPOP) work. It makes the sampled graph mimic original graph so that relational classification algorithms can be evaluated meaningfully. [14] observes that there are multiple graphs on OSN except for static friendships and propose a random walk sampling on the union of those graphs, which exhibits more rapid mixing. [34] improves RDS and MHRW by preventing backtracking. [7] presents a spectral sparsifier with only $n/\varepsilon^2$ edges. However, it takes $O(n^4)$ for the construction which renders it impractical for many problems. [68] studies network reconstruction from multiple SBS samples and node level attributes. It also give numerical evaluation of community detection algorithm performance on the sampled (reconstructed) and original graph.